\documentstyle[aps,prb,epsf]{revtex}

\newcommand{\Tk}{{\em $T_{K}$}}
\newcommand{\Tv}{{\em $T_{v}$}}

\newcommand{\Ef}{{\em $E_{f}$}}

\newcommand{\nft}{{\em $n_{f}(T)$}}

\newcommand{\wn}{{$cm^{-1}$}}

\newcommand{\sigo}{{\em $\sigma_{1} (\omega)$}}

\newcommand{\w}{{\em $\omega$}}

\newcommand{\iocm}{{\em $\Omega^{-1} cm^{-1}$}}

\newcommand{\sigdc}{{\em $\sigma_{d.c.}$}}
\newcommand{\ybin}{YbInCu$_{4}$}

\newcommand{\ybb}{YbB$_{12}$}

\begin{document}
\twocolumn[\hsize\textwidth\columnwidth\hsize\csname@twocolumnfalse%
\endcsname
\bibliographystyle{unsrt}
\title{Observation of a new excitation in the mixed-valent state of \ybin\ }
\author{S. R. Garner, Y.W. Rodriguez, and Z. Schlesinger}
\address{Department of Physics, University of California\\
Santa Cruz, California 95064}
\author{B. Bucher}
\address{Department of Physics, ITR\\
Oberseestrasse 10;  8640 Rapperswil   
Switzerland}
\author{Z. Fisk}
\address{Department of Physics and NHMFL, Florida State University\\
Talahassee, Florida 32310}
\author{J. L. Sarrao}
\address{Los Alamos National Laboratory, Los Alamos\\
New Mexico, 84545}
\date{September 19, 1999}
\maketitle
\begin{abstract}
Infrared measurements are used to obtain conductivity as a function of temperature and 
frequency in \ybin, which exhibits an isostructural transition to a mixed-valent state at 
$T_v \simeq 42 K.$ 
In addition to a gradual loss of spectral weight with decreasing temperature extending up to 1.5 eV,
sharp resonances appear in the mixed-valent state at 0 and 0.25 eV . These features may 
be key to understanding both \ybin\ and the nature of the mixed-valent Kondo state.
\end{abstract}
]
The presence of local moments in metallic systems is 
associated with a variety of interesting
phenomena, including the Kondo effect, heavy-fermion physics and mixed-valence\cite{hewson}.  
In rare cases, an isostructural
first-order transition at which a discontinuous change
in valence and volume accompanies an abrupt disappearance
of the local moment is observed\cite{lawrence0,allen2,nowik1,nowik2,sarraop}.  
This moment disappearance can be described in terms of the formation
of a mixed-valent Kondo-singlet state, in which
the f-level moment is compensated  
due to a Kondo-like screening by conduction electrons\cite{hewson}.
The transition to such a state provides an exceptional opportunity to probe 
a range of fundamental phenomena, including moment compensation, Kondo singlet
formation, and mixed-valence.

The prototypical example of a volume/valence transition is the $\gamma - \alpha$
transition\cite{lawrence0} of Ce, in which a valence change from about 3 to 3.2
occurs in concert with a volume reduction of about 15\% at \Tv $\sim$200 K .
According to the Kondo-volume-collapse model\cite{allen2}, 
the reduced lattice constant in the low temperature phase
is associated with an increase in hybridization between
local moment and conduction electron states. This results in an enhanced Kondo energy,
which drives the transition to a Kondo-singlet ground state.
The energy reduction associated with the formation of the
singlet ground state justifies the loss of entropy associated
with the disappearance of the local-moment degrees-of-freedom.
Technical difficulties associated with an intermediate phase\cite{lawrence0} 
make it very difficult to study the intrinsic physics of this transition in Ce.

\ybin\ also exhibits a
transition to a mixed-valent Kondo-singlet ground state, which is isostructural
to its high-temperature local moment state.
In this compound the intrinsic physics is more accessible, as there is 
no intervening phase, and the transition occurs at $T_v \simeq42$ K
at ambient pressure\cite{kindler,lawrence2,sarrao2,immer,lawrence3}.
At this transition the Yb valence decreases from $\sim$3 to $\sim$2.85,
and the local moment vanishes.
The volume change is of 
opposite sign to that of Ce-- a difference consonant with 
the observation that Yb has one hole in the f-level, whereas Ce 
has one {\em f} electron; however 
the magnitude of the volume change in \ybin\ ($\simeq0.5\%$) is too small
to provide a basis for an increase in hybridization
that would drive the transition\cite{cornelius97,figueroa98,sarraot}.  \ybin\ is thus a very 
interesting system, with a transition from a magnetic state to a mixed-valent
ground state that is not well understood.

In this letter, we focus on changes in the infrared 
conductivity of \ybin\ associated with the transition into the mixed-valent state.  
The abrupt increase of the Kondo scale below \Tv\ may allow us to identify key features
of the Kondo state, and thus shed light on fundamental phenomena of
Anderson lattice systems. This work is complementary to previous optical work
which addressed the relationship between spectral features and band-structure
calculations\cite{marab,galli,cont1,cont2}. 

At the transition two resonances appear.  The first is a Drude-like peak centered at zero 
frequency which is qualitatively
similar to low-temperature behavior seen in certain cerium compounds (c.f. ref. 21).  
The second is a resonance at $\sim0.25$ eV, which is present only in the mixed-valent state. 
We discuss the interpretation of these resonances as intra- and inter-band excitations
of coherent 
Kondo-state quasiparticles, respectively, for which the substantial increase of \Tk\
at \Tv\ is critical.  
\begin{figure}[htbp]
\epsfxsize=3.8in
\centerline{\epsffile{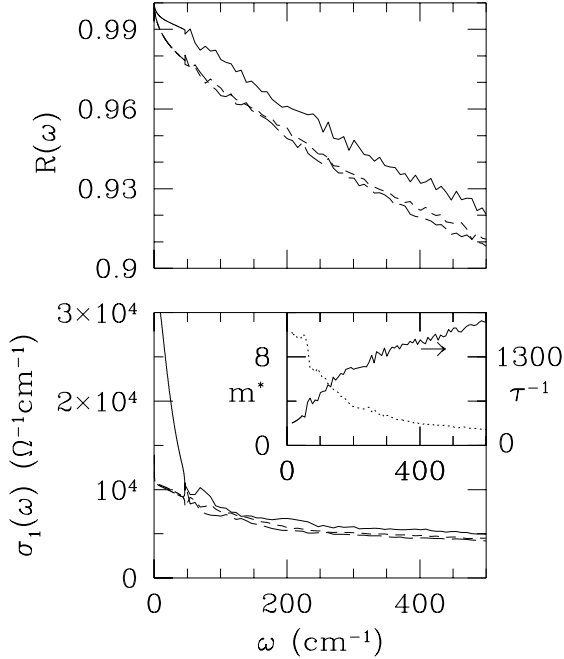}}
\caption{
The reflectivity and the real part of the conductivity at low frequency
are shown for \ybin\ at T= 250 K (long dashes), 55 K (shorter dashes) and 20 K (solid).  
The low-frequency resonance at 20 K is associated with the intraband excitations of a long-lived
quasi-particle (Kondo) resonance.  The inset shows the scattering rate (solid), $\tau^{-1},$  in $cm^{-1}$,
and effective mass enhancement (dotted), $m^{*}$, vs frequency at 20 K.
}
\label{fig1}
\end{figure}

\noindent 
Also of interest are 
spectral weight changes extending beyond 1 eV,
which may have implications regarding the energy, time and length scales associated 
with moment compensation and Kondo singlet formation.

The samples used in these experiments are high-quality single crystals grown from 
an In-Cu flux\cite{sarrao2}.
For these samples a sharp transition occurs at about 42 K in the absence
of strain.  At the transition the volume increases by about 0.5 \% as the sample
is cooled, and the
susceptibility and resistivity drop abruptly by an order of magnitude.
Thermal cycling tends to induce strain in the samples, which can broaden
the transition and move it to higher temperature\cite{sarrao2}.
Infrared and optical measurements are performed using a combination
of Fourier transform and grating spectrometers to cover the range from
50 to 50,000 \wn .  In these measurements we have gone to great efforts to measure in
all ranges before going through the transition to avoid disorder effects
influencing the infrared data significantly.
The conductivity as a function of frequency is obtained from a Kramers-Kronig
transform of the reflectivity data.  
For the purpose of performing this transform, the measured
reflectivity is extended from 50,000 to 200,000 \wn\ as a constant, and above
that it is made to decrease like 1/$\omega^2$.  At low frequency a Hagen-Rubens
termination is attached to the data.  
In the region of the actual data,
the conductivity is insensitive to the details of these terminations.

Figure 1 shows the reflectivity and the real part of the conductivity
in the low-frequency region in which a narrow Drude-like peak appears at low temperature.
Above \Tv\ the conductivity is suppressed and only weakly dependent on frequency,
due to the the strong scattering of the conduction electrons by the dense magnetic
``impurities'' (local moments).  Below \Tv\ this scattering is suppressed, the d.c.
resistivity decreases abruptly\cite{sarrao2} and a narrow resonance appears in \sigo ,
as shown in figure 1.  Extrapolated values of \sigo\ to \w = 0  of about
\sigdc $\simeq$10,000 \iocm\ above \Tv\ , and $\simeq40,000$ to 80,000 \iocm\ below \Tv\ are 
consistent with d.c. resistivity measurements\cite{kindler}.

\begin{figure}[htbp]
\epsfxsize=3.8in
\centerline{\epsffile{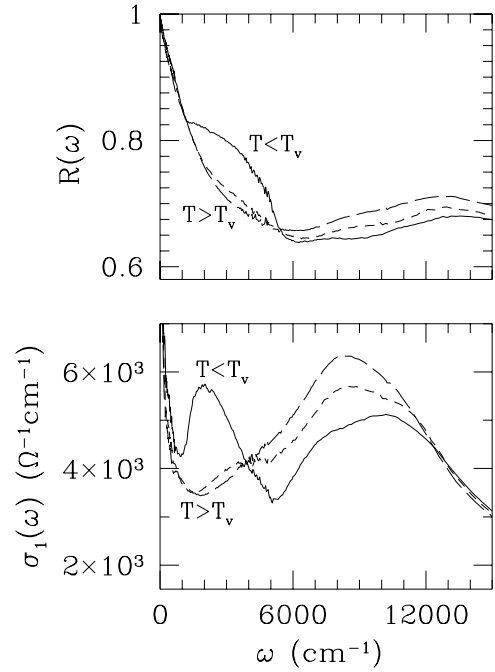}}
\caption{
The reflectivity and the real part of the conductivity at low frequency
are shown for \ybin\ at T= 250 K (long dashes), 55 K (shorter dashes) and 20 K (solid).  
Gradual reduction of spectral weight with cooling occurs in the vicinity of 8,000 \wn\ (1 eV).
A well-defined resonance appears at 2,000 \wn\ (1/4 eV) in the low temperature 
Kondo state.
}
\label{fig2}
\end{figure}

One can view this low-temperature behavior in terms of a frequency dependent scattering 
rate and effective mass\cite{allen0,pink1,webb}, as shown in figure 1b (inset).
At 20 K the scattering rate rises rapidly between about 25 and 200 \wn\
exhibiting a change in slope in the vicinity of 200 \wn , which is comparable
to the Kondo scale ($\simeq$280 \wn\ ) of the low-T state of \ybin .  
The effective mass enhancement increases with decreasing \w\  over the same range 
and approaches an asymptotic value of about $m^{\ast}\!\approx\!10$ at low frequency.
These low temperature quantities exhibit a crossover from a low energy
regime where the compensated moments are ineffective scatterers,
to a high energy regime in which conduction electrons 
are strongly scattered by uncompensated moments.
This reflects the evolution of the dynamics from that of dressed, heavy 
quasiparticles to that of the undressed band-like carriers, which is fundamental to
systems with a local moment resonance not too far from the chemical potential. 

Figure 2 shows reflectivity and conductivity to higher frequency (12,000 \wn ). 
These data show the persistence of significant temperature dependence 
to very high frequency (compared to T or \Tk ) in \ybin .
For example, between about 5,000 to 12,000 \wn\ 
\sigo\ decreases substantially as T is reduced both above and below \Tv\ .
In addition, a prominent resonance appears in the mixed-valent (low-T) state
near 2,000 \wn . 
The spectral sharpness and abrupt
appearance of this feature at $\omega\!\simeq\!2,000$ \wn\ ($\simeq\!1/4 eV$) as a function of temperature are striking.

Figure 3 shows spectral weight, which is the indefinite integral of \sigo\ ,
$n(\omega) = \frac{m}{\pi e^2}\int_0^\omega \sigma_1 (\omega^{\prime})\,\mathrm{d}\omega^{\prime}$,
as a function of frequency.
In this figure (and figure 2) we see that there is a net loss of spectral weight
as the temperature is lowered from 250 K to 55 K.  The loss amounts to about 10\% of the strength of 
the broad mode centered around 9,000 \wn , and corresponds to $\sim1$ carrier/Yb
atom with the reasonable assumption of a band mass of 3 (times the free-electron mass).  
Since spectral weight is ultimately
conserved (if one integrates to high enough frequency\cite{wooten}), 
these data imply that it must be displaced to
still higher frequency (above 16,000 \wn $\simeq2$ eV) as T is reduced from 250 to 55 K.  
Recent theoretical work\cite{rozenberg,freericks2} has explored possible
origins of such high energy spectral weight shifts 
(involving energies vastly larger than $K_{B}T$ and $K_{B}T_K$) 
in strongly correlated systems. 

The coalescence of the 20 and 55 K curves at the high frequency end of figure 3  
indicates that the increase in spectral weight associated with
the appearance below \Tv\ of the resonance at $\sim\!2000$ \wn\ is balanced by a general 
reduction of \sigo\ up to $\sim\!12,000$ \wn .
The displaced spectral weight corresponds to about 1.5 carriers/Yb.

Although the spectral weight of the very narrow low temperature resonance 
at $\omega=0$ (figure 1) is quite small,
it is significant to the correspondence between the infrared data and Hall
effect data for \ybin .  Hall effect, when corrected for skew scattering, reflects 
an increase from about 0.7 carriers/Yb above \Tv , to a much higher value
($\sim\!4$ carriers/Yb) below the transition\cite{cornelius97}.
Above \Tv , the low frequency rise of the conductivity (figure 2) can be fit 
with a broad ($\sim\!700$ \wn ) resonance with a strength which corresponds to 
about one carrier per Yb, which is roughly consistent with the high temperature Hall data.  
Below \Tv\ a much sharper ($\sim\!25$ \wn ) additional peak appears in \sigo\
at $\omega\sim0$, as seen in figure 1b.
With the inclusion of the frequency dependent, low-temperature mass enhancement of
$m^{*}\!\approx\!10$ (figure 1b, inset), 
this narrow peak represents an additional 2.5 carriers/Yb,
consistent with the substantial increase in carrier density inferred from the low T Hall effect data.

\begin{figure}[htbp]
\epsfxsize=2.8in
\centerline{\epsffile{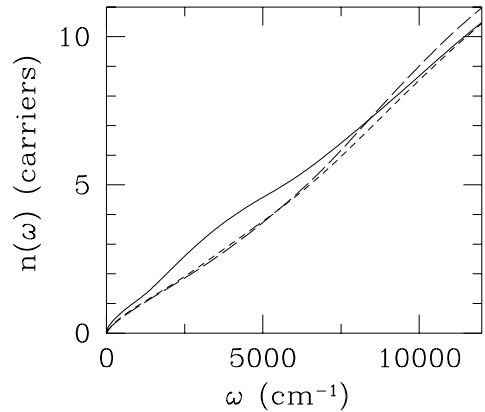}}
\caption{
Spectral weight,
the integral of \sigo\ as a function of frequency from 0 to \w , is shown as a function of
\w\ at T= 250 K (long dashes), 55 K (shorter dashes) and 20 K (solid).  
The units for the vertical axis, expressed in terms of carriers per Yb atom, 
are established by the assumption of a band mass of 4.
}
\label{fig3}
\end{figure}

Both the starting Hamiltonian and the mechanism that drives the transition to the 
mixed-valent state remain areas of active research for \ybin\ .  
With regard to the mechanism, it has been argued that the 
lattice expansion is too small to explain the large change in Kondo temperature 
(from $\sim$25 K to 400 K) at the transition\cite{sarraot}.
The Falikov-Kimball model is capable of producing a quasi Hubbard-like first-order transition, 
and may be relevant to high-temperature properties of \ybin ,
however it ignores hybridization,
which is certainly important in the low-T state\cite{freericks}.
In the mixed-valent state, where the Kondo scale is large,
the dynamics of the Periodic Anderson Model (PAM) are expected to be relevant.
Within the PAM context, the 1/4 eV excitation can be associated with a quasiparticle
interband transition involving Kondo resonance states 
near \Ef\ \cite{coleman,jarrell}.
The abrupt change of \Tk\ at the transition and the abrupt appearance of the resonance 
are consistent with this interpretation.  The study of \ybin\ , with its first-order transition
at which \Tk\ increases by an order of magnitude, thus appears to allow the first clear
identification of this fundamental excitation.

The energy scale for this interband excitation involving the
dynamically generated quasi-particle states at \Ef\ (the Kondo resonance) 
is expected to be\cite{coleman}  $\sim\!\sqrt{T_K B}$.
Since $T_K \simeq \tilde{V}^{2} / B$ this provides a measure of the renormalized 
hybridization, $\tilde{V}$.
Using the value $\tilde{V}\simeq1/4$ eV from our infrared data, along with
$T_K\simeq400$ K ($\simeq35$ meV), implies a bandwidth of $B\simeq1.8$ eV, which is
reasonable.  One can estimate the hybridization broadening, $\Gamma$,
using its relationship to $\tilde{V}$, to be $\Gamma\simeq0.25$ eV. 
Further, one can use $\Gamma$ in NCA formulae\cite{bickers87} involving
\nft\ along with $L_{III}$ edge measurements of valence\cite{sarrao99}
to infer that the f-level is about 0.5 eV away from the chemical potential.
These values are quite reasonable for this mixed valent system.

The observation that the growth of the resonance at $\simeq$1/4 eV comes from
a redistribution of spectral weight from essentially the entire range below 1.5 eV
(comparable to the bandwidth)
may have implications for questions related to exhaustion and the time scales relevant to 
screening in Kondo lattice systems\cite{jarrell,millis}.
Does it suggest that conduction electrons
further than $K_B T_K$ from the chemical potential are significantly 
involved in screening in the Kondo lattice?
Further work can be expected to address such questions.
It is also intriguing to note that
an excitation of similar frequency is present in \ybb\ \cite{okamura}, 
for which $T_K \simeq 300$ K, and that related features may also be present in 
spectra from mixed-valent Ce compounds\cite{bucher3}.

In summary, \ybin\ is of interest because of the rarity
of valence transitions, a lack of understanding of their underlying mechanism,
and due to the opportunity to observe the effect of dramatic changes 
of $T / T_K$ on physical properties.	
We observe high-energy spectral weight changes, which may be relevant
to the mechanism, and the abrupt appearance of a sharp excitation 
near 1/4 eV, present only in the high $T_K / T$ state,  
which is interpreted as the Kondo-quasiparticle interband excitation. 

Acknowledgements:  The authors acknowledge valuable conversations with
J. W. Allen, D. L. Cox, P. Coleman, J. K. Freericks, D. H. Lee and A. P. Young,
and technical assistance from Todd Lorey, Sonya Hoobler, Jason Hancock and Petar Kostic.
Work at UCSC supported by the NSF through grant\# DMR-97-05442.
Work at Los Alamos is performed under the auspice of the U.S. Dept. of
Energy.  NHMFL is supported by the NSF and the state of Florida.  ZF and
JLS also acknowledge partial support from the NSF under grant \# DMR-9501529.

\bibliography{zs-short,ybcu,valence,kondo99}
\end{document}